\documentclass[aps,pre]{revtex4}

\usepackage{graphicx}
\usepackage[utf8]{inputenc}
\usepackage[english]{babel}
\usepackage{amsmath}
\usepackage{enumerate}
\usepackage{epstopdf}
\usepackage{amssymb}

\begin{document}
\title{Langevin equations from experimental data: the case of rotational diffusion in granular media} 
\author{Marco Baldovin $^{1}$, Andrea Puglisi $^{2}$ and  Angelo Vulpiani $^{1,3}$}  

\affiliation{%
$^{1}$ Dipartimento di Fisica, ``Sapienza'' Universit\`a di Roma, p.le A. Moro 2, 00185 Roma, Italy;\\
$^{2}$ CNR-ISC and Dipartimento di Fisica, Sapienza Universit\`a di Roma, p.le A. Moro 2, 00185 Roma, Italy;\\
$^{3}$ Centro Linceo Interdisciplinare ``B. Segre'', Accademia dei Lincei, via della Lungara 10, 00165 Rome, Italy} 

\date{\today}

\begin{abstract}
 A model has two main aims: predicting the behavior
of a physical system and understanding its nature, that is how it
works, at some desired level of abstraction. A promising recent
approach to model building consists in {\em deriving} a Langevin-type
stochastic equation from a time series of empirical data. Even if the
 protocol is based upon the introduction of drift and
diffusion terms in stochastic differential equations, its
implementation involves subtle conceptual problems and, most
importantly, requires some prior theoretical knowledge about the
system. Here we apply this approach to the data obtained in a rotational granular
diffusion experiment, showing the power of this method
and the theoretical issues behind its limits. A crucial point emerged
in the dense liquid regime, where the data reveal a complex
multiscale scenario with at least one fast and one slow
variable. Identifying the latter is a major problem within the
Langevin derivation procedure and led us to introduce innovative ideas
for its solution.
\end{abstract}


\maketitle
\section{Introduction}

The Langevin equation is surely one of the pillars of non
equilibrium statistical mechanics~\cite{zwanzig}.  Such a stochastic process has been
introduced more than one century ago by Langevin in his seminal paper
on the Brownian motion of colloidal particles in a fluid.  In a
nutshell the basic idea is the following: in a system with some slow
variables it is possible to model the dynamics of these observables with
an effective stochastic equation containing a systematic drift and a
noisy term.


The study of Brownian motion played a crucial role to establish, in a
conclusive way, the physical validity of the atomic
hypothesis~\cite{ld2014}. In his celebrated work Langevin had been
able to write the Brownian evolution law with a deep intuition and a
clever combination of macroscopic and microscopic ingredients (namely
the Stokes law and energy equipartition). This work had been one of
the starting points of the mathematical theory of continuous
stochastic processes, namely stochastic differential equations, which
are basically a generalisation of the Langevin
equation~\cite{vanKampen2007,gardiner1985handbook,R89}. In the
following we will use ``Langevin equation'' with the loose meaning of
stochastic differential equation, an identification broadly accepted in the literature.

Unfortunately a stochastic differential equation can be derived from a 
microscopic description with a systematic approach just in few cases.  One 
important example is the diffusion of a big heavy intruder in a diluted gas of 
light particles: the kinetic theory allows to determine the stochastic 
differential equation ruling the evolution of the velocity of the heavy particle 
in terms of the microscopic parameters~\cite{cecconi2007transport}.

Another system where it is possible to build the Langevin equation with an 
analytical approach is a large harmonic chain containing $N$ particles, one of 
which much heavier than the others, in the limit $N \gg 1$~\cite{rubin60}.

As far as we know, the origin of the (few) successes in the derivation of a 
stochastic differential equation with a non phenomenological approach can be always related to 
the high dilution of the system or to its linear character.

On the other hand, typically, it is necessary to adopt a more
pragmatic attitude combining mathematics (if possible), intuition,
suggestions from the data, and a preliminary understanding of the
system under investigation. The present paper aims at applying a
recently developed method to derive a Langevin equation from a time
series of data~\cite{peinke11,baldolang}. This method relies upon the
basic textbook definitions of drift and diffusion coefficients of a
stochastic differential equation, but in its applications it has been
refined under many aspects, which are not only technical but also
conceptual. Some difficulties are related to the coexistence of two
properties of the Langevin equation, i.e. continuity and
Markovianity. The most relevant conceptual issue is that of
determining the proper variables for a complete Markovian description,
which is not trivial when the available data are constituted by a
single variable time series: in the theory of dynamical systems the
so-called embedding theorem (due to Takens) provides a procedure - in
principle of general validity - to reconstruct the correct phase
space~\cite{Sauer1991}, once one assumes that the system is deterministic.
Unfortunately such a
procedure in many specific cases, including the one discussed here, is
not useful and must be replaced by new ideas. Methods based on a
stochastic approach are widely used in order to model and analyse
complex features in a general setting, e.g. $1/f$ noise, see for instance~\cite{m1,m2,m3};
our aim here is to find effective Markov models, with quantitative
determination of parameters, for a specific system.

The present paper is organized as follows.  In the Materials and
Methods section we first discuss the procedure of Langevin derivation
from data, with its practical and conceptual difficulties and, second,
we revise the experimental setup (rotational diffusion in diluted and
dense granular gases) and some previous phenomenological models
adopted to understand such experiment.  In the Results section we
apply the method to the data, in particular in a dilute and a dense
case, which are extremely different cases.  In the Conclusion section
the reader can find general and conclusive remarks.

\section{Materials and methods}

\subsection{Langevin equation from data}

Let us present a direct approach to build a Langevin equation from data.
 As far as we know, in spite of its (relatively) easy use, there has been
just few attempts in such a direction. 
Likely it is due to the fact that, although the 
method can appear quite obvious and easy at a first glance,
there are some rather severe difficulties both at conceptual 
and technical level.
\\
Let us assume that we know that the slow ``good'' variables are the component of a vector ${\bf X}(t) \in R^N$.
In the following we will see that  this is a rather subtle point, and the choice of the proper ${\bf X}$ is
not easy at all.
Let us also suppose that we have measured, e.g. in an experiment or a simulation, a long time series of these variables,  $ \{ {\bf X}(t) \}$.
With these assumptions, one can determine the $N$ Langevin equations ($n \in [1,N]$)
\begin{equation}
\label{le}
{d X_n \over dt}=F_n({\bf X})+\sqrt{2D_n({\bf X})} \eta_n,
\end{equation}
where $\eta_n(t)$ are white noises i.e. Gaussian processes with $\langle \eta_n(t)\rangle=0$ and $\langle \eta_n(t) \eta_{n'}(t')\rangle=\delta_{nn'} \delta(t-t')$,
following the definitions found in textbooks on stochastic processes~\cite{vanKampen2007,gardiner1985handbook,R89},
that is in terms of the statistical features of
$$
\Delta X_n(\Delta t)=X_n(t+\Delta t)-X_n(t).
$$ 
In fact the drifts and diffusion coefficients are given by the following formula:
\begin{subequations}
\label{d&d}
\begin{align}
\label{drift}
F_n({\bf X})&= \lim_{\Delta t \to 0} { 1 \over \Delta t} 
\langle \Delta X_n(\Delta t) | {\bf X}(t)={\bf X} \rangle \\
\label{diffusion}
D_n({\bf X})&= \lim_{\Delta t \to 0} { 1 \over 2\Delta t} 
\langle {(\Delta X_n(\Delta t)-F_n({\bf X})\Delta t)}^2 | {\bf X}(t)={\bf X} \rangle 
\end{align}
\end{subequations}
It is easy to realize, see Ref.~\cite{baldolang}, that the limit
$\Delta t \to 0$ must be considered in a proper physical sense,
i.e. smaller than typical time, but not too small. Sometimes the term
``Langevin equation'' is used with different meanings, here it
indicates Eq.~\eqref{le} where ${\bf F}$ is not necessary linear and
the noise may be multiplicative, i.e. $D$ can depend upon ${\bf X}$.

The practical procedure to extract the $\{ F_n \}$ and $\{ D_{n}\}$ from
data is not trivial; however it is not the most difficult
problem one has to face: the main conceptual trouble is indeed given by the
absence a general method for choosing the ``right'' variables,
an aspect that is too often overlooked. For instance, Onsager
and Machlup are explicit in raising the question \cite{onsager}:
\begin{quote}
 How do you know you have enough variables, for [the system] to be Markovian?
\end{quote}  
Similarly, Shang-Keng Ma expresses a caveat of central importance \cite{ma85}:
\begin{quote}
 The hidden worry of thermodynamics is: we do not know how many
  coordinates or forces are necessary to completely specify an
  equilibrium state.
\end{quote}
Usually the ``proper variables'' are unknown
and in the building of the model one can use the time series of
just one observable $\{ U(t) \}$, or a few ones.  Such problem is quite
similar to the phase space reconstruction in dynamical systems.  There
are no automatic protocols for this choice, and typically to obtain
some good results it is necessary to possess the expertise and/or
intuition about the problem under investigation. For a general
discussion on the difficulties to build models from data,
see~\cite{noientr}.

Sometimes mathematics can help to anticipate that a certain set of
variables is not adequate as Markovian model. For instance, see
Ref.~\cite{baldolang}, in the case of a single scalar variable, the
shape of the correlation function may be already sufficient to exclude
that such variable is an equilibrium Markov process, so that it
becomes necessary to look for a new set of variables.

\subsection{Experimental setup and phenomenological models}

Granular materials apparently share many properties with condensed
``molecular'' matter~\cite{JNB96,BPS15}, but such similitudes hide a
crucial difference: grains, being macroscopic, dissipate energy
through friction (in enduring contacts or rapid collisions). For this
reason equilibrium statistical physics may only suggest qualitative
ideas for fluidized steady states and dramatically fails in the
extreme case of static or quasi-static regimes.  The liquid state of
granular matter, which is in the middle between fast ``granular
gases'' and slow ``granular glasses'', feels stronger the need for a
coherent theoretical framework: continuum descriptions for dense flows
lack first-principle constitutive relations~\cite{andreotti,luding}, so that transport coefficients can only be measured in molecular dynamics simulations~\cite{khain},
while kinetic theories (e.g. mode-coupling) must be carefully adapted
to take into account some fundamental peculiarities, such as
dissipation and inertia~\cite{zippelius}. An important insight is
provided by experiments, where such a liquid state is obtained through
some mild shaking of the container~\cite{danna,hecke2,camille}. In the
setup described below our focus is on regimes where the longest
relaxation time is reasonably smaller than the total experimental
time, so that the system can be said to be in a {\em steady state}. In
a word we are not interested, here, in the solid or glassy
states~\cite{dyre,barrat,bouchaud}.

\begin{figure}[h!]
\includegraphics[width=5.2in,clip=true]{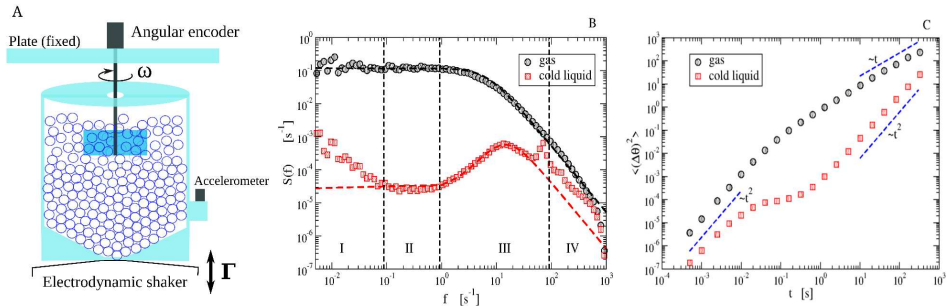}
\caption{{\bf Experimental results}
  A: Sketch of the experiment reported in~\cite{camille}. B: Experimental
  data of the VPDS for the gas case and the ``cold liquid'' case,
  together with predictions (dashed lines) from the incomplete
  model, Eq~\eqref{dhc}. C: Experimental data of the MSD for both cases,
  together with dashed lines useful as guides for the
  eye.\label{Fig1} }
\end{figure}

A recent experimental study~\cite{camille} has offered a new picture
for dense granular flows in a wide range of time-scales, from
$10^{-3}$ s up to $10^3$ s and more, revealing an unexpectedly rich
scenario. In the experimental setup, sketched in
Fig~\ref{Fig1}A the ``impurity'' was constituted by an
immersed blade who could rotate around a fixed vertical axis under
the kicks from the grain of a vibrofluidized granular medium. The
dynamics of the angular velocity $\omega(t)$ of the blade and its
absolute angular position $\theta(t)=\int_0^t ds \omega(s)$, was
studied in different regimes of density and intensity of vibration. In
Fig~\ref{Fig1}B, the velocity power density spectrum (VPDS),
\begin{equation}
S(f)=\frac{1}{2\pi t_{TOT}}\lvert\int_0^{t_{TOT}} \omega(t) e^{i  (2\pi f) t}dt \rvert^2 ,
\end{equation}
is presented and its salient features are
highlighted in two opposite limits, which are the gas and the cold
liquid. We remind that the VPDS is the Fourier transform of the
velocity autocorrelation function and that its $f \to 0^+$
limit is the self-diffusion coefficient, i.e. $D_\infty=\pi \lim_{f\to
  0^+} S(f)$. We also recall that relations exist, under certain
approximations, between the VPDS and the intermediate scattering
function which - in liquids - is typically accessed through neutron
scattering experiments \cite{rahman}.

In the gas limit (low packing fraction and high energy per grain) the
probe velocity autocorrelation is close to a simple exponential
decay $\sim e^{-t/\tau_{gas}}$, ruled by a single relaxation time
$\tau_{gas}$: in this limit the VPDS takes the form of a Lorentzian
\begin{equation}
S(f)=\frac{T}{\pi \gamma}\frac{1}{1+(2\pi If/\gamma)^2}.
\end{equation}
In the - roughly
speaking - opposite limit, that of a ``cold liquid'' (high packing
fraction $\gtrsim 30-35\%$\, and low energy per grain), the observed
VPDS strongly deviates from the Lorentzian. Ignoring a mechanical
resonance due to the mounting plate at $\sim 10^2 Hz$, it displays
four different regions: at high frequency (region IV) it decays with a
negative power law equal or smaller than $2$; in region III it shows a
smooth parabolic maximum (centered near $\sim 10 Hz$), reminiscent of a
harmonic confinement (``cage'') typical of molecular and granular
liquids~\cite{cavagna,marty,reis2007,zippelius}; in region II it
stabilizes on a short plateau, which suggests a loss of memory (as in
the plateau of the Lorentzian which marks the onset of normal
diffusion); finally region I, perhaps the most surprising one, shows a
diverging $S(f)$ for $f \to 0^+$, signaling a problem with the
finiteness of the self-diffusion coefficient $D_\infty$. A few longer
experiments (12 hours) were conducted, showing a slow crossover toward
a new higher plateau at very low frequencies. The study of the mean
squared displacement (MSD), see Fig~\ref{Fig1}C confirmed that
the four regions of the cold liquid case correspond, respectively, to
short-time ballistic (free) motion (IV), dynamical arrest due to
caging (III), later relaxation of the cage (II) and ``final''
superdiffusive behavior (I), very rarely observed in previous works on
granular systems~\cite{barrat,bouchaud,behringer,roux}.  A universal
scenario for anomalous diffusion is lacking~\cite{klages}, but
certainly it is the signal of an enduring memory. A family of
phenomenological models for anomalous diffusion includes fractional
Fokker-Planck equations~\cite{klafter}, where an immediate physical
interpretation is not always at hand, or phenomenological continuous
time random walk model for the {\em velocity}~\cite{castiglione}, with
a power-law-decaying distribution of persistency times (see
Supplemental Materials of~\cite{camille}).  Interestingly, simpler
models - e.g. {\em linear} Langevin equations - can offer an even more
complete insight in the many observed phenomena.

In~\cite{camille} a first model was proposed to account for the caging phenomenon, i.e. regions
II-III-IV of the VPDS in the cold liquid limit, inspired by the  Itinerant
Oscillator model for molecular liquids~\cite{io1,io2,vollm}; it is described by the
following stochastic equations of motion:
\begin{subequations} \label{dhc}
\begin{align}
I\dot{\omega}(t)&=-\gamma\omega(t)-k[\theta(t)-\theta_0(t)]+\sqrt{2 \gamma T}\eta(t)\\
\dot{\theta}(t)&=\omega(t) \\
\dot{\theta}_0(t)&=\sqrt{2D_0}\eta_0(t)
\end{align}
\end{subequations}
where $\eta(t)$ and $\eta_0(t)$ are independent white normal Gaussian
noises (unitary variance) and the angles $\theta(t)$ and $\theta_0(t)$ are considered not bounded. The model represents the diffusion of a
particle in a harmonic potential with ``stiffness'' $k$ and unfixed
minimum located at $\theta_0(t)$, under the effect of a thermal bath
at temperature $T$ and relaxation time $I/\gamma$. The harmonic
potential, representing the cage created by the confining effect of
the dense granular host fluid, is not fixed but moves, as
$\theta_0(t)$ behaves as Brownian motion with diffusivity
$D_0$. Motivation for this model is twofold: 1) it reproduces the main
features of the VPDS, i.e. short time fast relaxation (region IV), an
elastic resonance at intermediate times (region III) and a plateau
revealing loss of memory at larger times (region II); 2) in the dilute
limit (when $k \to 0$) there are analytical arguments for it~\cite{cecconi2007transport,bromo},
3) at intermediate densities a series of studies showed that memory
effects (coming from correlated collisions) are well described by a
similar coupling with an additional degree of freedom characterized by
slower relaxation time-scales~\cite{sar10}.  The VPDS of the above
model is fully calculated in~\cite{camille}, fairly reproducing the VPDS in regions II-IV  (see dashed lines in Fig~\ref{Fig1}B) in which $k/I \sim (2\pi \cdot 10)^2 \sim 4 \cdot 10^3 Hz^2$.

In order to overcome the strong disagreement between the previous
model and observations in region I (super-diffusion), a new model was
introduced in~\cite{lasanta}:
\begin{subequations} \label{model}
\begin{align}  
I \dot{\omega}(t)&=-\gamma \omega(t) -k[ \theta(t) - \theta_{0}(t)]+ \sqrt{2\gamma T}\eta(t)\\ 
I_{0}\dot{\omega}_{0}(t)&=-\gamma_{0} \omega_{0}(t) +k[ \theta(t) - \theta_{0}(t)]+ \sqrt{2\gamma_0 T_0}\eta_{0}(t) \label{model2}\\
\dot{\theta}(t)&=\omega(t)\\
\dot{\theta}_0(t)&=\omega_0(t)
\end{align}
\end{subequations}
In equations~\eqref{model} the angular velocity of the probe feels 
two different forces both related to collisions: one part is without memory and is described by the
 $-\gamma\omega+\eta(t)$  contribution, the
second part takes the form $-k[ \theta(t) - \theta_{0}(t)]$ and
therefore depends upon the past history of $\omega(t)$ and
$\omega_0(t)$.
As before $\theta_0(t)$ should be
viewed as a collective degree of freedom representing the preferential
point of the blade with respect to some granular {\em cage}. The {\em
  cage} slowly changes its configuration and favors the blade's drift
at later times. This model however replaced the overdamped dynamics of
$\theta_0(t)$ in Eq~\eqref{dhc}, introducing the crucial effect of
cage inertia $I_0$. For the sake of symmetry a reciprocal effect of
the blade upon the granular material was included, an ingredient which
is likely to be negligible in view of the large value of
$I_0$. In~\cite{lasanta} the introduction of cage inertia was
motivated only by analogy: it was reasonable, when looking for an
ingredient reproducing almost ballistic superdiffusion, to imagine
that the cage (which, over long time-scales, dominates also the probe's dynamics) is doing long
ballistic drifts sustained by its large inertia. Discreteness and
finiteness of the granular material, which in the experiments is made
of a few thousands grains, makes random but persistent (also called
``secular''~\cite{PADCC14}) drifts possible.

The linearity of the model allowed to solve it analytically, with a
closed formula for many statistical properties, for instance the VPDS
and, semi-analytically, also for the MSD. After some controlled
procedure of fitting for the many parameters it was possible to find a
perfect reproduction of the full dynamics in regions I-IV, including
super-diffusion at long times, with a value of $I_0$ of the same order
of the moment of inertia of the granular medium surrounding the probe in the
experiment~\cite{lasanta}.

\section{Results}

In the previous section we have discussed how experimental data presented 
in~\cite{camille} can be described by simplified stochastic models, whose 
parameters can be found by fitting dynamical observables as the VPDS or the MSD. 
In the following we will face the problem from a different, complementary, point 
of view: instead of checking the accordance of the measurements with a 
predeterminate theoretical model (based on physical arguments), we will try to 
get important information on the model itself directly from data, by enforcing 
the extrapolation protocol discussed above.

If the studied variable (namely, the angular velocity of the probe as a function
of time) is a continuous Markov process, the procedure should be able to
``automatically'' find the best functional form for the corresponding Langevin 
Equation (provided that the sampling frequency is high enough). In the
following we will examine a case, the gas limit, in which this scheme can be 
applied quite straightforwardly; in the cold-liquid limit, on the contrary, 
some additional considerations from physics will be needed -- and not always 
sufficient -- in order to get a satisfactory description.

\subsection{Gas limit}

Let us consider first a ``dilute gas'' case: the container is filled with 350 
grains, corresponding to a packing fraction of $\phi=5\%$; the shaking intensity 
is $\Gamma=\ddot{z}/g=39.8$, where $\ddot{z}$ stands for the vertical acceleration and 
$g=9.81 m/s^2$ is the gravitational acceleration. The measuring set-up records the 
angular position $\theta(t)$ of the blade with a sampling rate of $f_s=2000 Hz$, 
so that we can compute the angular velocity $\omega(t)=\dot{\theta}(t)$ with a 
temporal resolution of $\Delta t_{min}=1/f=0.5 ms$. Analyzing a long time 
series (1 hour) of data, we would like to infer the parameters $F(\omega)$ and 
$D(\omega)$ of Eq~\eqref{le}.

\begin{figure}[h!]
\includegraphics[width=5.2in,clip=true]{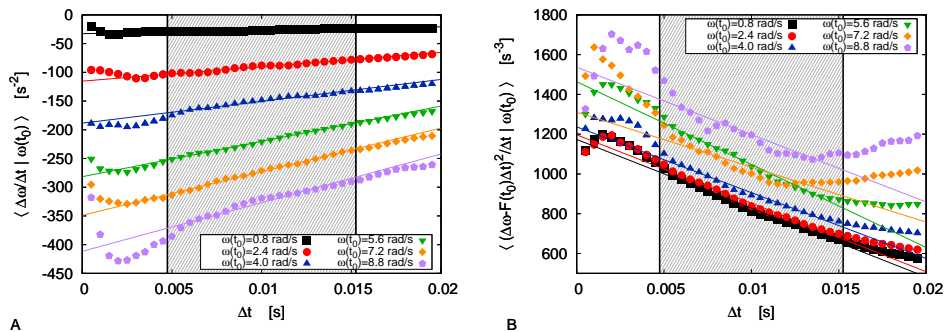}
\caption{{\bf Gas limit: extrapolation}
Extrapolation of the limits on the r.h.s. of Eq~\eqref{d&d}, for 
several values of $\omega(t_0)$, in order to compute drift (panel A) and 
diffusivity (panel B) in the gas limit. Each linear fit (solid lines) has been 
computed - by means of a classical least-square method - considering only the data in the shaded range, within the vertical lines.
\label{Fig2}}
\end{figure}

In Fig~\ref{Fig2} we plot the average quantities that appear on the 
r.h.s. of Eq~\eqref{d&d}, for several values of the time interval 
$\Delta t$. As discussed in \cite{peinke11} and recalled in the previous section, when 
studying data series resulting from deterministic physical processes, the 
Markovian approximation can be considered true only at suitable time scales, 
namely for $\tau_{ME}\ll\Delta t\ll\tau$, where $\tau_{ME}$ is the 
Markov-Einstein time and $\tau$ is a characteristic time for the autocorrelation 
function of the considered process. As a consequence, the limits on the r.h.s. 
of Eq~\eqref{d&d} should be evaluated as extrapolations of the trend presented 
by data in a suitable time-scale range ($\Delta t \in [0.005, 0.015] s$ in our 
case, shaded region in Fig~\ref{Fig2}).
 We can perform a linear extrapolation using the least-square method: the vertical
 intercept of the resulting graph is our guess for the limit $\Delta t \rightarrow 0$.
 In order to evaluate a confidence interval for such value, one could estimate the uncertainty
 of each point of the graph and then consider the error propagation on the vertical intercept;
 however, since the data are not independent, this method is expected to underestimate the
 uncertainty. A safer way to compute the confidence interval is the ``jackknife
 method'' \cite{jack}: here we have divided our sampled data into $n=100$ blocks, then 
 we have repeated the analysis $n$ times, discarding one block at each turn,
 and we have computed the confidence interval from the distribution of the resulting $n$ different expected values.
 
\begin{figure}[h!]
\centering
\includegraphics[width=3.2in,clip=true]{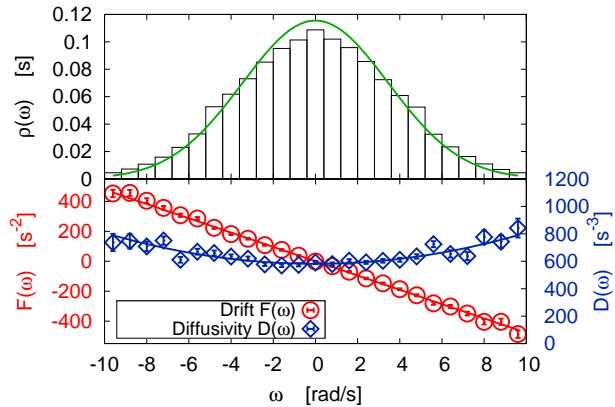}
\caption{{\bf Gas limit: PDF, drift and diffusivity} Top: Probability
  distribution (PDF) of $\omega$ in the gas limit (the green solid
  line is the prediction of the derived model). Bottom: Reconstructed
  drift (red circles) and diffusivity (blue diamonds) in the gas limit
  have been respectively fitted with a linear and a parabolic
  function.
\label{Fig3}}
\end{figure}

Taking the $\Delta t \rightarrow 0$ limit of the extrapolated linear
trends, we have an estimate for the drift coefficient $F(\omega)$ and
for the diffusivity $D(\omega)$: as it is shown in Fig~\ref{Fig3} (bottom),
the former has a linear dependence $F(\omega)=-A\omega$ on the angular velocity, while
the latter can be approximated as $D(\omega)=D_1+D_2\omega^2$.
 Of course, our procedure gives more accurate results when the angular velocity is smaller,
 i.e. when a bigger volume of data is available for the averages (see Fig~\ref{Fig3} (top)).
Let us notice that the quadratic corrections to the diffusivity are only relevant when
$|\omega|$ is quite large, i.e. when our estimate is less reliable
because of the low volume of data. For this reason, in the following we neglect
such corrections and apply the constant approximation
$D(\omega)=D$. Model~\eqref{le} reduces then to the well known
Ornstein-Uhlenbeck process \cite{gardiner1985handbook}, so that all
interesting physical observables can be computed analytically.

\begin{figure}[h!]
\includegraphics[width=5.2in,clip=true]{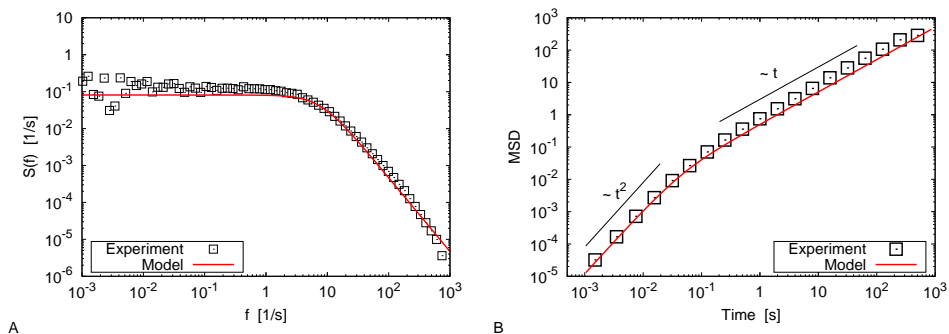}
\caption{{\bf Gas limit: observables}
Velocity power density spectrum (panel A) and mean square displacement 
(panel B) in the gas limit. Experimental data (black squares) are compared with 
the reconstructed model (red lines). Black lines are guides for the eyes. 
\label{Fig4}}
\end{figure}

In Fig~\ref{Fig3} (top) we observe a fair agreement between the
predicted stationary probability distribution of $\omega$ and the
experimental one. Table~\ref{Tab1} summarizes the expected values for the
parameters of the model, and the corresponding uncertainties.
\begin{table}[h!]
 \centering
 \begin{tabular}{|c|c|}
 \hline
 Parameter & Value\\
 \hline
 $F$ & $(47.82\pm 0.42) s^{-1}$\\
 $D$ & $(581.6\pm 5.8) s^{-3}$\\
 \hline
\end{tabular}
\caption{{\bf Gas limit: parameters} Expected values and uncertainties
for the parameters of the reconstructed model in the gas limit.}
 \label{Tab1}
\end{table}
In Fig~\ref{Fig4} we compare the experimental VPDS
and MSD with the theoretical ones for the reconstructed
Ornstein-Uhlenbeck process, finding a good agreement. The gas limit
can be fairly approximated by this model, as already discussed in
\cite{camille}. It is useful to recall that the characteristic time of
the Ornstein-Uhlenbeck process (i.e. the decay of the velocity
autocorrelation) is proportional to the mean free time between
particle-blade collisions and in certain conditions can be
quantitatively predicted~\cite{GPT13}. We stress that if one considers
also the quadratic corrections and performs numerical simulations, the
outcomes are almost identical, at least in this case.

\subsection{Cold liquid limit}

In the following we analyze a regime which is somehow ``opposite'' to the gas 
limit seen above: in this case we consider $N=2600$ beads and a shaking 
intensity $\Gamma=39.8$; the packing fraction is $\phi=36\%$. Again, $f=2000 Hz$ 
and the experiment has a duration of 1 hour.

As already understood in \cite{camille, lasanta}, in this case the rich 
phenomenology of the system cannot be described by a single-variable approach, 
since the dynamics of the granular matter involves at least two clearly separate 
time scales. Before enforcing the extrapolation procedure, we should be able to 
identify a ``fast'' variable and a ``slow'' one in order to understand how the 
model depends on them.

\begin{figure}[h!]
\includegraphics[width=5.2in,clip=true]{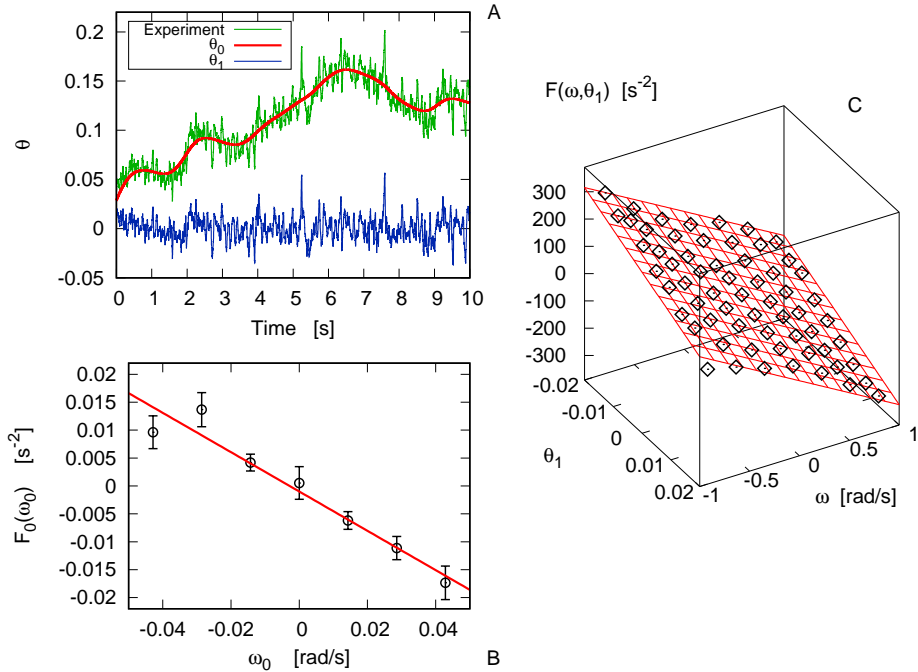}
\caption{{\bf Cold liquid: slow and fast components}
A: Decomposing the original signal (green line) into the sum of the 
averaged angular position $\theta_0$ defined by Eq~\eqref{avetheta} (red line) 
and the ``fast'' variable $\theta_1=\theta-\theta_0$ (blue line). B: Drift 
coefficient of $\omega_0$. C: Drift coefficient of $\omega$, as a function of 
$\theta_1$ and $\omega_0$. \label{Fig5}}
\end{figure}

A quite straightforward way to define a variable that describes the slow 
behavior of the probe is to consider a running average with a Gaussian window 
function:
\begin{equation}
 \label{avetheta}
 \theta_0(t)=\frac{1}{\sqrt{2\pi}\sigma}\int dt'\,e^{-\frac{(t'-t)^2}{\sigma^2}}\theta(t'-t)\,.
\end{equation}
The fast component can be found, of course, as $\theta_1(t)=\theta(t)-\theta_0(t)$ (see
Fig~\ref{Fig5}A).
The value of the characteristic time $\sigma$ (here $\sigma=0.3 s$) is suggested by the shape of $S(f)$, see
Fig.~\ref{Fig1}: we need to filter out the features in regions III and
IV, but we also demand that the interesting dynamics in region I is reproduced
by the new variable; taking $2 \sigma \simeq O(1) s$ seems therefore a legitimate choice.
In the following we will show that varying the parameter does not affect the results of our analysis significatively,
as long as $\sigma$ is chosen to be of the same order of magnitude.
Of course, any other choice for the kernel in Eq~\eqref{avetheta} could be made,
provided that it canceled the fast oscillations of $\theta(t)$
(i.e. provided that it had a Fourier transform decaying fast
enough).

Let us notice that, following the physical
interpretation proposed in \cite{lasanta}, $\theta_0$ can be seen as
the center of mass of the itinerant ``cage'' at every time, while
$\theta_1$ has the meaning of the angular distance between $\theta_0$
and the probe itself.

First, we can use the extrapolation analysis seen before in order to determine a 
proper Langevin equation for the slow variable $\omega_0=\dot{\theta}_0$. In this case the 
significative $\Delta t$ range can be found at a much slower time scale (namely, 
$\Delta t \in [1.5, 6] s$): as a consequence, the volume of available data 
considerably shrinks, but it is still possible to estimate the dependence of the 
drift term on the slow angular velocity. In particular one finds 
(Fig~\ref{Fig5}B) that $F(\omega_0)= -A_0\omega_0$ is an acceptable 
approximation. As in the previous case, we approximate the diffusivity term with 
a constant, $D_0(\omega)=B_0$, neglecting the deviations for large $|\omega_0|$.

We are left with the problem of finding a model for the observed variable 
$\omega(t)$. In Fig~\ref{Fig5}C we show that the drift coefficient of 
$\omega$ depends significantly not only on $\omega$ itself, but also on 
$\theta_1=\theta-\theta_0$. A linear function of both arguments, 
$F(\omega,\theta_1)=- A_1 \omega-A_2 \theta_1 $, turns out to provide a fair 
description of the data.  Again we consider a constant value for the 
diffusivity, $D(\theta_1, \omega)=B$.

The reconstructed model reads as follows:
 \begin{subequations}
  \label{model5}
   \begin{align}
    \dot{\omega}(t)&= - A_1 \omega(t) -A_2 [\theta(t)-\theta_0(t)]+ \sqrt{2B}\eta(t)\\
    \dot{\omega_0}(t)&=-A_0 \omega_0(t) + \sqrt{2B_0}\eta_0(t)\\
    \dot{\theta}(t)&=\omega(t)\\
\dot{\theta}_0(t)&=\omega_0(t)
  \end{align}
\end{subequations}
where $\eta(t)$ and $\eta_0(t)$ are Gaussian noises with unitary
variance. Such system is a particular limit of the model guessed in
\cite{lasanta}, where the term $k[\theta(t)-\theta_0(t)]$ is
negligible with respect to the other terms in Eq.~\eqref{model}b:
indeed with the present definition, $\omega_0$ evolves with a slow
dynamics that does not admit fluctuations on the fast time scale of
$\theta_1$, so that such term is necessarily negligible.

\begin{figure}[h!]
\includegraphics[width=5.2in,clip=true]{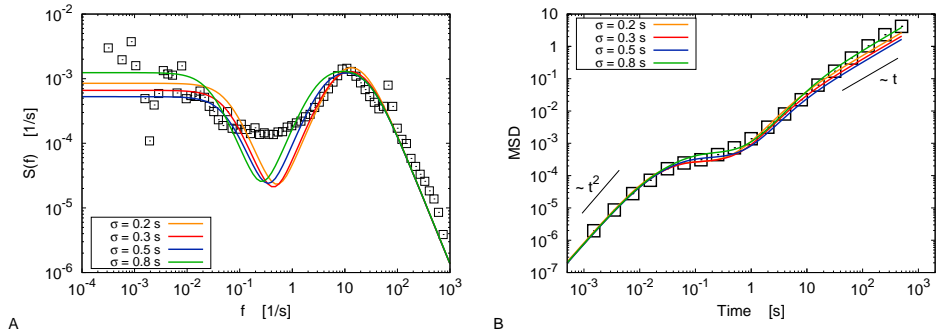}
\caption{{\bf Cold liquid: observables}
Velocity power density spectrum (panel A) and mean square displacement 
(panel B) in the cold liquid case with $\Gamma=39.8$. Experimental data (black 
squares) are compared with the reconstructed models (coloured lines) for several values of
the smoothing parameter $\sigma$. Black lines are 
guides for the eyes. \label{Fig6}}
\end{figure}

The power density spectrum of $\omega(t)$ for model~\eqref{model5} can be determined analytically:
\begin{equation}
  S(\hat{f})=\frac{1}{\pi}\frac{A_2^2 B_0 + A_0^2 B \hat{f}^2 + B \hat{f}^4}{ A_0^2 A_2^2 + [A_2^2+ A_0^2 A_1^2 - 2 A_0^2 A_2^2]\hat{f}^2 + [A_1^2 - 2 A_2 + A_0^2]\hat{f}^4 + \hat{f}^6}
\end{equation}
where $\hat{f}=2\pi f$. Once $S(\hat{f})$ is known, the MSD can be found as
\begin{equation}
  \langle [\Delta \theta(t)]^2\rangle=\int_0^t dt' \int_0^t dt'' \langle \omega(t') \omega(t) \rangle = 2 \int_0^t dt' (t-t') C_{\omega \omega}(t')
\end{equation}
where $C_{\omega \omega}(t)$, the autocorrelation function of $\omega(t)$, is 
the Fourier anti-transform of $S(\hat{f})$. In Fig~\ref{Fig6} we compare the 
above analytical expressions to the experimental data.

The VPDS shows a fair agreement in the high-frequency regime, $f\gg1 Hz$, and in 
the low-frequency one $f\ll1 Hz$; in the intermediate range (region II in the 
notation of \cite{camille}) there is a clear discrepancy between the model and 
the experimental data, maybe due to decorrelations of the slow variable that are 
not caught by the model. However, we stress that the difference concerns a 
frequency range which is almost inessential for the dynamical properties of the 
system, whose characteristic frequencies lay in regions I and III of the 
spectrum (see Fig~\ref{Fig1}B): this is completely evident when considering 
the MSD evolution (Fig~\ref{Fig6}B), that is very well reproduced by the 
model despite the discrepancy on $S(\hat{f})$.
Let us note that changing $\sigma$ by a factor 4, from $0.2 s$ to $0.8 s$, does not affect
the results of our analysis in a significative way.

Finally, let us consider an experiment with $N=2600$, $\Gamma=26.8$, 
$\phi=45\%$: even if the number of beads is the same as in the previous case, 
the lower shaking intensity entails that the accessible volume for the beads is 
lower, i.e. the actual packing fraction increases. The system is therefore in a 
``more concentrate'' state. The duration of the experiment has also been raised 
(12 hours), and we have chosen $\sigma=0.3 s$ for the analysis.

\begin{figure}[h!]
\includegraphics[width=5.2in,clip=true]{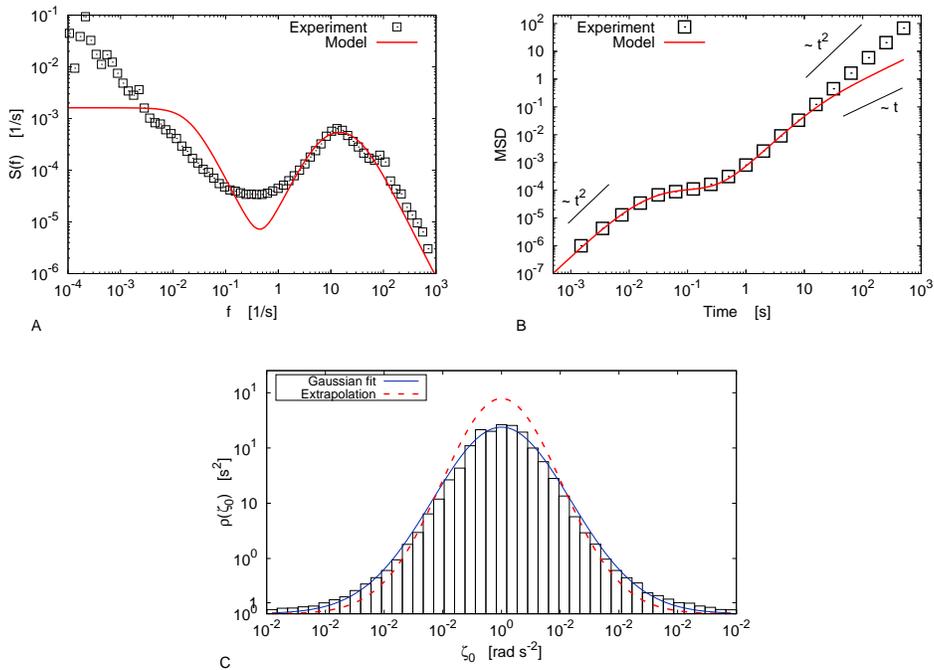}
\caption{{\bf Cold liquid: limits of the method}
A: Velocity power density spectrum in the cold liquid limit with 
$\Gamma=26.8$. B: Mean square displacement in the same case. C: Distribution of 
the ``noise'' $\zeta_0$ defined by Eq~\eqref{noise} (with $\Delta_t=0.6 s$), 
Gaussian fit (blue line) and comparison with the reconstructed model (dashed red 
line).\label{Fig7}}
\end{figure}

Fig~\ref{Fig7}A and Fig~\ref{Fig7}B show the VPDS and the MSD in 
this case, compared to those that can be inferred through the method discussed 
above. Even if the high-frequency regime is well reproduced by the model, our 
description fails on the long-time scale; in particular, the MSD of the 
reconstructed model shows a linear dependence on time for $t\gtrsim 50 s$, while 
the experimental one is still quadratic on that scale.

The failure of the method could be ascribed to the choice of Gaussian noise for 
the slow variable: if the evolution of $\omega_0$ was ruled by a Lévy process, 
an alternative analysis should be considered \cite{siegert01}. Let us evaluate, 
\textit{a posteriori}, the noise of the slow variable as:
\begin{equation}
 \label{noise}
 \zeta_0(t)=\frac{\omega_0(t+\Delta t)-\omega_0(t)}{\Delta t}-F_0(\omega_0(t))\,.
\end{equation}
Fig~\ref{Fig7}C shows that the distribution of $\zeta_0$ can be fairly 
approximated with a Gaussian; furthermore, the amplitude of the noise is very 
close to that of the reconstructed model. Hence, we can guess that the 
hypothesis of Gaussian noise is quite reasonable, and the discrepancy between 
the experimental data and the inferred model could need a different explanation.

 Table~\ref{Tab2} summarizes the expected values for the
parameters of model~\eqref{model5}, and the corresponding uncertainties,
for the two considered cases in the cold liquid limit.
Also in this case the confidence intervals have been computed using the jackknife method
on $n$ blocks of sampled data: we have chosen $n=100$ for the fast variables, $n=10$ for the slow ones.
\begin{table}[h!]
 \centering
 \begin{tabular}{|c|c|c|}
 \hline
 Parameter & $\Gamma=39.8$ & $\Gamma=26.8$\\
 \hline
 $A_1\; [s^{-1}]$ & $ 200.3\pm 4.4 $ & $ 252.0\pm 6.3 $\\
 $A_2\; [s^{-2}]$ & $ 5.76\cdot 10^3\pm 2.4 \cdot 10^2 $ & $ 8.55 \cdot 10^3\pm 4.5 \cdot 10^2 $\\
 $B\; [s^{-3}]$ & $ 161.1\pm 3.9 $ & $ 107.3\pm 3.1 $\\
 $A_0\; [s^{-1}]$ & $ 0.352\pm 0.034 $ & $ 0.1317\pm 4.9 \cdot 10^{-3} $\\
 $B_0\; [s^{-3}]$ & $ 2.43 \cdot 10^{-4}\pm 2.1 \cdot 10^{-5} $ & $ 8.82 \cdot 10^{-5}\pm 2.7 \cdot 10^{-6} $\\
 \hline
\end{tabular}
\caption{{\bf Cold liquid: parameters} Expected values and uncertainties
for the parameters of model~\eqref{model5} in the cold liquid limit.}
 \label{Tab2}
\end{table}



\section{Conclusion}

Using data from an accurate experiment on the rotational diffusion in
granular media we derived a Langevin equation which is able to
describe the main dynamical statistical features of the system.
\\ Let us stress that, in the building of the model, part of the
protocol is quite standard. On the other hand there are rather subtle
conceptual aspects which cannot be ignored.  In the dilute case it is
enough to use a single variable whose dynamics is well described by a
linear Langevin equation.  Much more difficult is the dense case where
it is not obvious, as already stressed in the past by many eminent
scientists~\cite{onsager,ma85}, the set of variables which is ruled by
a Markov process.  \\ At a first glance, it seems natural to follow
the approach of Takens for the phase space
reconstruction~\cite{Sauer1991}.  The basic idea is: from the
knowledge of a time series $\{ u(t) \}$ one starts with the variable
$\{ u(t) \}$ itself, if this choice is not appropriate one can try
with $\{ u(t), du(t)/dt \}$, then $\{ u(t), du(t)/dt, d^2u(t)/dt^2\}$,
and so on. Once the dimension of the phase space has been fixed one
can try to find the evolution equation, certainly with many
practical problems to solve, see~\cite{KS}. In our system one faces an
additional important obstacle. Consider a series $\{ u(t) \}$ with a
characteristic time $\tau_c$, surely the variables $du(t)/dt$,
$d^2u(t)/dt^2$ etc.  have characteristic times which cannot be larger
that $\tau_c$, therefore a protocol based on the Takens' approach cannot succeed in systems with
a multiscale structure, i.e. with other relevant variables whose time
scales are much longer than $\tau_c$.  \\ Actually in our case in
order to find a good model we have been forced to go in a direction
which is the opposite of the Takens method; i.e.  to identify a slow
variable obtained with a convolution, which is somehow antithetical to
a derivative.

\section*{Acknowledgments}
We thank A. Gnoli and A. Plati for useful discussions.


\begin{thebibliography}{50}

\bibitem{zwanzig}
Zwanzig R.
\newblock Nonequilibrium Statistical Mechanics.
\newblock Oxford University Press; 2001.

\bibitem{ld2014}
Vulpiani A, Cecconi F, Cencini M, Puglisi A, Vergni D, editors.
\newblock {Large Deviations in Physics}.
\newblock Springer; 2014.

\bibitem{vanKampen2007}
van Kampen NG.
\newblock {Stochastic Processes in Physics and Chemistry}.
\newblock Elsevier; 2007.

\bibitem{gardiner1985handbook}
Gardiner CW.
\newblock {Handbook of stochastic methods}.
\newblock Springer Berlin; 1985.

\bibitem{R89}
Risken H.
\newblock The Fokker-Planck equation: Methods of solution and applications.
\newblock Berlin: Springer- {V}erlag; 1989.

\bibitem{cecconi2007transport}
Cecconi F, Cencini M, Vulpiani A.
\newblock {Transport properties of chaotic and non-chaotic many particle
  systems}.
\newblock Journal of Statistical Mechanics: Theory and Experiment.
  2007;2007:P12001.

\bibitem{rubin60}
Rubin RJ.
\newblock {Statistical dynamics of simple cubic lattices. Model for the study
  of Brownian motion}.
\newblock Journal of Mathematical Physics. 1960;1(4):309.

\bibitem{peinke11}
Friedrich R, Peinke J, Sahimi M, Tabar MRR.
\newblock Approaching complexity by stochastic methods: From biological systems
  to turbulence.
\newblock Phys Rep. 2011;506:87.

\bibitem{baldolang}
Baldovin M, Puglisi A, Vulpiani A.
\newblock Langevin equation in systems with also negative temperatures.
\newblock J Stat Mech. 2018;2018(4):043207.
\newblock doi:{10.1088/1742-5468/aab687}.

\bibitem{Sauer1991}
Sauer T, Yorke JA, Casdagli M.
\newblock Embedology.
\newblock J Stat Phys. 1991;65(3-4):579--616.
\newblock doi:{10.1007/BF01053745}.

\bibitem{m1}
Mao X, Shang P, Wang J, Ma Y.
\newblock Characterizing time series by extended complexity-entropy curves
  based on Tsallis, R{\'e}nyi, and power spectral entropy.
\newblock Chaos: An Interdisciplinary Journal of Nonlinear Science.
  2018;28(11):113106.

\bibitem{m2}
Leibovich N, Dechant A, Lutz E, Barkai E.
\newblock Aging Wiener-Khinchin theorem and critical exponents of $1/f^\beta$
  noise.
\newblock Physical Review E. 2016;94(5):052130.

\bibitem{m3}
Rodr{\'\i}guez MA.
\newblock Complete spectral scaling of time series: Towards a classification of
  1/f noise.
\newblock Physical Review E. 2014;90(4):042122.

\bibitem{onsager}
Onsager L, Machlup S.
\newblock Fluctuations and Irreversible Processes.
\newblock Phys Rev. 1953;91:1505.

\bibitem{ma85}
Ma SK.
\newblock {Statistical Mechanics}.
\newblock World Scientific Publishing; 1985.

\bibitem{noientr}
Baldovin M, Cecconi F, Cencini M, Puglisi A, Vulpiani A.
\newblock The Role of Data in Model Building and Prediction: A Survey Through
  Examples.
\newblock Entropy. 2018;20:807.

\bibitem{JNB96}
Jaeger HM, Nagel SR, Behringer RP.
\newblock The physics of granular materials.
\newblock Physics Today. 1996;49(April):32.

\bibitem{BPS15}
Baldassarri A, Puglisi A, Sarracino A.
\newblock Coarsening in granular systems.
\newblock C R Physique. 2015;16:291.

\bibitem{andreotti}
Andreotti B, Forterre Y, Pouliquen O.
\newblock Granular Media. Between Fluid and Solid.
\newblock Cambridge University Press; 2013.

\bibitem{luding}
{S  Luding}.
\newblock {Towards dense, realistic granular media in 2D }.
\newblock Nonlinearity. 2009;22(22):R101.
\newblock doi:{10.1088/0951-7715/22/12/R01}.

\bibitem{khain}
Khain E.
\newblock Thermal conductivity at the high-density limit and the levitating
  granular cluster.
\newblock Phys Rev E. 2018;98:012903.
\newblock doi:{10.1103/PhysRevE.98.012903}.

\bibitem{zippelius}
Fiege A, Aspelmeier T, Zippelius A.
\newblock Long-Time Tails and Cage Effect in Driven Granular Fluids.
\newblock Phys Rev Lett. 2009;102:098001.

\bibitem{danna}
D'Anna G, Mayor P, Barrat A, Loreto V, Nori F.
\newblock Observing brownian motion in vibration-fluidized granular matter.
\newblock Nature. 2003;424:909.

\bibitem{hecke2}
Wortel GH, Dijksman JA, van Hecke M.
\newblock Rheology of weakly vibrated granular media.
\newblock Phys Rev E. 2014;89:012202.

\bibitem{camille}
Scalliet C, Gnoli A, Puglisi A, Vulpiani A.
\newblock Cages and anomalous diffusion in vibrated dense granular media.
\newblock Phys Rev Lett. 2015;114:198001.

\bibitem{dyre}
Dyre JC.
\newblock The glass transition and elastic models of glass-forming liquids.
\newblock Rev Mod Phys. 2006;78:953.

\bibitem{barrat}
Heussinger C, Berthier L, Barrat JL.
\newblock Superdiffusive, heterogeneous, and collective particle motion near
  the fluid-solid transition in athermal disordered materials.
\newblock Europhys Lett. 2010;90:20005.

\bibitem{bouchaud}
Lechenault F, Dauchot O, Biroli G, Bouchaud JP.
\newblock Critical scaling and thereogeneous superdiffusion across the
  jamming/rigidity transition of a granular gas.
\newblock Europhys Lett. 2008;83:46003.

\bibitem{rahman}
Rahman A.
\newblock Correlations in the motion of Atoms in Liquid Argon.
\newblock Physical Review. 1964;136:A405.

\bibitem{cavagna}
Cavagna A.
\newblock Supercooled liquids for pedestrians.
\newblock Phys Rep. 2009;476:51.

\bibitem{marty}
Marty G, Dauchot O.
\newblock Subdiffusion and Cage effect in a Sheared Granular Material.
\newblock Phys Rev Lett. 2005;94:015701.

\bibitem{reis2007}
Reis PM, Ingale RA, Shattuck MD.
\newblock Caging Dynamics in a Granular Fluid.
\newblock Phys Rev Lett. 2007;98:188301.

\bibitem{behringer}
Utter B, Behringer RP.
\newblock Self-diffusion in dense granular shear flows.
\newblock Phys Rev E. 2004;69:031308.

\bibitem{roux}
Radjai F, Roux S.
\newblock Turbulentlike Fluctuations in Quasistatic Flow of Granular Media.
\newblock Phys Rev Lett. 2002;89:064302.

\bibitem{klages}
Klages R, Radons G, Sokolov IM, editors.
\newblock Anomalous transport.
\newblock Wiley\&Sons; 2008.

\bibitem{klafter}
Klafter J, Lim SC, Metzler R, editors.
\newblock Fractional Dynamics: Recent Advances.
\newblock World Scientific; 2012.

\bibitem{castiglione}
Andersen KH, Castiglione P, Mazzino A, Vulpiani A.
\newblock Simple stochastic models showing strong anomalous diffusion.
\newblock Eur Phys J B. 2000;18:447.

\bibitem{io1}
Sears VF.
\newblock The itinerant oscillator model of liquids.
\newblock Proc Phys Soc. 1965;86:953.

\bibitem{io2}
Coffey W, Kalmykov YP, Waldron JT.
\newblock The Langevin Equation: With Applications in Physics, Chemistry, and
  Electrical Engineering.
\newblock World Scientific; 1996.

\bibitem{vollm}
Vollmer HD.
\newblock Correlations in the motion of Atoms in Liquid Argon.
\newblock Z Physik. 1979;33:103--109.

\bibitem{bromo}
Sarracino A, Villamaina D, Costantini G, Puglisi A.
\newblock Granular Brownian motion.
\newblock J Stat Mech. 2010;2010:P04013.

\bibitem{sar10}
Sarracino A, Villamaina D, Gradenigo G, Puglisi A.
\newblock Irreversible dynamics of a massive intruder in dense granular fluids.
\newblock Europhys Lett. 2010;92:34001.

\bibitem{lasanta}
Lasanta A, Puglisi A.
\newblock An itinerant oscillator model with cage inertia for mesorheological
  granular experiments.
\newblock J Chem Phys. 2015;143:064511.

\bibitem{PADCC14}
Pons A, Amon A, Darnige T, Crassous J, Cl\'ement E.
\newblock Mechanical fluctuations suppress the threshold of soft-glassy solids
  : the secular drift scenario.
\newblock Phys Rev E. 2015;92:020201(R).

\bibitem{jack}
Flyvbjerg H.
\newblock Error estimates on averages of correlated data.
\newblock In: Kert{\'e}sz J, Kondor I, editors. Advances in Computer
  Simulation. Berlin, Heidelberg: Springer Berlin Heidelberg; 1998. p. 88--103.

\bibitem{GPT13}
Gnoli A, Puglisi A, Touchette H.
\newblock Granular Brownian motion with dry friction.
\newblock Europhys Lett. 2013;102:14002.

\bibitem{siegert01}
Siegert S, Friedrich R.
\newblock Modeling of nonlinear L\'evy processes by data analysis.
\newblock Phys Rev E. 2001;64:041107.

\bibitem{KS}
Kantz H, Schreiber T.
\newblock Nonlinear time series analysis. vol.~7.
\newblock Cambridge University Press; 2004.

\end{thebibliography}

%
%
%
%
%
%
%

\end{document}